\documentclass[multphys,vecphys]{svmult}

\usepackage{makeidx}         
\usepackage{graphicx}        
\usepackage{multicol}        
\usepackage[bottom]{footmisc}

\begin{document}

\title*{Observations of 51 Ophiuchi with MIDI at the VLTI}
 \author{C. Gil\inst{1}\and F. Malbet\inst{2} 
         \and M. Sch{\"o}ller\inst{1}
         \and O. Chesneau\inst{3} 
          \and Ch. Leinert\inst{4} 
                          }
\institute{European Southern Observatory,
              Casilla 19001, Santiago 19, Chile
              \texttt{cgil@eso.org, mschoell@eso.org}
         \and
             Laboratoire d'Astrophysique de l'Observatoire de Grenoble, BP 53 38041 Grenoble Cedex 9, France
             \texttt{Fabien.Malbet@obs.ujf-grenoble.fr}
          \and
           Observatoire de la C{\^o}te d'Azur, CNRS UMR 6203, Avenue Copernic, Grasse, France 
              \texttt{Olivier.Chesneau@obs-azur.fr}
              \and
              Max-Planck-Institut fur Astronomie, Knigstuhl 17, 69117 Heidelberg, Germany         \texttt{leinert@mpia-hd.mpg.de}}

%
%
\maketitle

\begin{abstract}{We present interferometric observations of the Be star 51 Ophiuchi. These observations were obtained during the science demonstration phase of the MIDI instrument at the Very Large Telescope Interferometer (VLTI). Using MIDI, a Michelson 2 beam combiner that operates at the N band (8 to 13 $\mu$m), we obtained for the first time observations of 51 Oph in the mid-infrared at high-angular resolution.

It is currently known that this object presents a circumstellar dust and gas disk that shows a very different composition from other Herbig Ae disks. The nature of the 51 Oph system is still a mystery to be solved. Does it have a companion? Is it a protoplanetary system? We still don't know. 
Observations with MIDI at the VLTI allowed us to reach high-angular resolution (20 mas).We have several uv points that allowed us to constrain the disk model. We have modeled 51 Oph visibilities and were able to constrain the size and geometry of the 51 Oph circumstellar disk.}
\end{abstract}

\section{Introduction}
\label{sec:1}

51 Ophiuchi is a Be (B9.5IIIe) star, located at 131 pc with a rotational velocity $v sini$ = 267 $\pm$ 5 km/s \cite{gil:Dunkin}. 51 Oph shows a large infrared excess, that was first noticed by Waters et al. 1988 \cite{gil:Waters}. This infrared excess was explained as being originated by the circumstellar dust around the star. 

The 51 Oph circumstellar gas and dust origin is still unknown. Since its infrared excess was noticed, this object has been observed in many different wavelengths, from the infrared to the ultraviolet. In 1993, Grady and Silvis \cite{gil:Grady} found the first evidence for the presence of dust by detecting a 10 $\mu$m silicate feature in emission. 51 Oph has been compared to  $\beta$ Pic due to the presence of a gas envelope with a variable column density, high density gas accreting to the star and the fact that this gas is also collisionally ionized. There was no detection of an extended disk at 18 $\mu$m, which suggested that the hot dust must be located in the close proximity of the star \cite{gil:Jayawardhana}. Van den Ancker et al. \cite{gil:vandenAncker} analyzed 51 Oph ISO archive data and suggested a few different scenarios for the nature of this system. The authors have studied the composition of the circumstellar gas and detected spectral features due to hot gas-phase molecules not typical of Ae/Be stars. They have suggested that 51 Oph might not be a young system but a highly evolved one instead. Nevertheless, the presence of all the circumstellar material can not be explained if the system is really an evolved one. The other scenario includes the presence of a companion in order to explain the hot gas observed. A more exotic explanation consists of assuming that the detected material is a result of a recent collision of two gas-rich planets or the accretion of a solid body as the star increases its size at the end of its main-sequence life \cite{gil:vandenAncker}.   

Roberge et al. \cite{gil:Roberge} observed 51 Oph circumstellar disk with the Far Ultraviolet Spectroscopic Explorer and found that the composition of the infalling gas is highly nonsolar. However, all studies suggest that the circumstellar matter is most likely in the form of a Keplerian disk rather than in a spherical shell. 
In 2004 the first MIDI observations of Herbig Ae/Be stars were published by Leinert et al. \cite{gil:Leinert}. The authors have studied a sample of 7 Herbig Ae/Be stars but could not fit a Dullemond et al. (2001) type disk model to the 51 Oph data like they did for the other stars. 

In this paper, we have used the data obtained by Leinert et al. \cite{gil:Leinert} on 51 Oph and constrained the size and geometry of the 51 Oph circumstellar disk by using a standard disk model. In section 2 the observations and data reduction are presented. The results are introduced and discussed in sections 3 and 4, and summarized in section 5.

\section{Observations and Data Reduction}
\label{sec:2}
51 Oph was observed during MIDI Science Demonstration Time (SDT) at the VLTI on June 15th and 16th 2003. MIDI is a 2 beam combiner that operates in the mid-infrared, 8 to 13 $\mu$m. The baseline used was the UT1-UT3 (102m), with a maximum full spatial resolution of 20 mas at 10 $\mu$m. The observations log is presented in Table~\ref{gil:log}. We have obtained 5 visibility points, that were dispersed at a resolution of 30 in the N band.

We have reduced the 51 Oph data using the IDL software written for MIDI.  The data reduction is described in detail in Leinert et al. (2004) \cite{gil:Leinert}. The data analysis and modeling was done with the interpreted programming language yorick.  

\begin{table}[h!] 
\centering
\label{gil:log}
\caption{Log of the observations \protect\cite{gil:log}}
\begin{footnotesize}
\begin{tabular}{c c c c c}
\hline
 UT Date &Projected      &  P.A.   & Hour Angle& Calibrators\\
         &  Baseline (m) & ($^\circ$)&         \\
\hline
15/Jun/03  &101.2 &23 & 3 &HD 168454, HD 168454, HD 167618 \\
15/Jun/03 & 101.4 &38 & 7 &HD 168454, HD 168454, HD 167618 \\
15/Jun/03 & 85.6  &45 & 8 &HD 168454, HD 168454, HD 167618 \\
16/Jun/03 & 98.8  &-7 & 0 &HD 165135, HD 152786, HD 165135  \\
16/Jun/03 & 99.6 &14  & 2 &HD 165135, HD 152786, HD 165135 \\
\hline 
\end{tabular}
\end{footnotesize}
\end{table}

\section{Results}
We have modeled the circumstellar disk by using a flat disk model with a temperature law proportional to $r^{-3/4}$ \cite{gil:Malbet}. The disk radial structure was derived from Lynden-Bell \& Pringle (1974). In this model, the disk is considered as infinitely thin and the flux is calculated through the sum of all the rings of radius $r$ and width $dr$.

In Table~\ref{gil:model} one can find a detailed list of the parameters used to fit the standard disk model to our data. The values for the stellar mass ($M_{\star}$), effective temperature ($T_{\rm eff}$) and atmospheric extinction (Av) were all taken from the literature \cite{gil:vandenAncker}.
 We have found a good fit for the data with a disk nearly edge-on, inclination of 88$^\circ$, position angle of 78$^\circ$ and an accretion rate of 7.$10^{-5}$ $M_{\odot}$/yr.

\begin{table}[t!]
\label{gil:model}
\begin{center}
\caption{Disk model parameters for 51 Oph.\protect\cite{gil:model}}
\begin{tabular}{|c|c|}
\hline
Parameter & Best fit Value \\
\hline
Distance & 131 pc\\
 $R_{\star}$ & 7 $R_{\odot}$\\
$M_{\star}$ & 3.8 $M_{\odot}$\\
$T_{\rm eff}$ & 10000 K\\
Av& 0.15\\
Accretion rate&  7.$10^{-5}$ $M_{\odot}$/yr\\
Disk outer radius& 7 AU\\
Disk inner radius& 0.55 AU\\
Inclination & 88$^\circ$\\
Position Angle & 78$^\circ$\\
\hline
\end{tabular}
\end{center}
\end{table}

\begin{figure}[h!]
 \label{gil:sed}
 \centering
 \includegraphics[height=7cm]{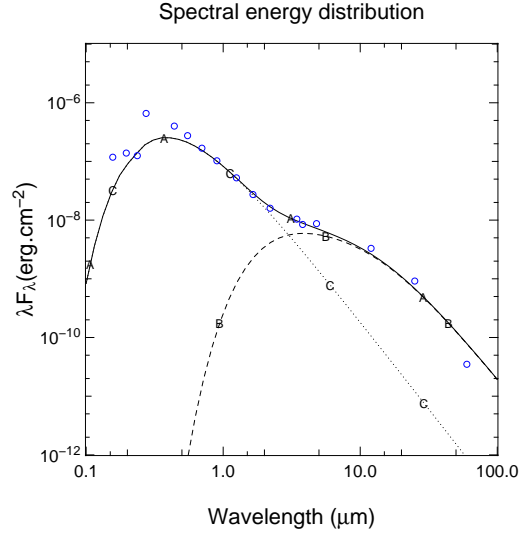}
 \caption{Spectral energy distribution for 51 Oph. The circles correspond to photometry measurements compiled from the literature (Waters et al. 1988). The dotted line is the stellar contribution, the dashed line is the disk contribution and the solid line corresponds to the total energy distribution (star+disk). Our standard disk model successfully reproduces the 51 Oph SED.\protect\cite{gil:sed}}
 \end{figure}

For  a radius of 7 $R_{\odot}$, we have found a good fit for the spectral energy distribution (SED), with an inner disk radius of 0.55 AU and an outer disk radius of 7 AU (see Fig.~\ref{gil:sed}). In this figure one can see the SED for 51 Oph well reproduced by a standard disk model. The photometry points used in these figure were taken from Waters et al. (1988) \cite{gil:Waters}.  

\begin{figure*}[h!]
\label{gil:uv}
 \centering 
\centerline{\includegraphics[height=6cm]{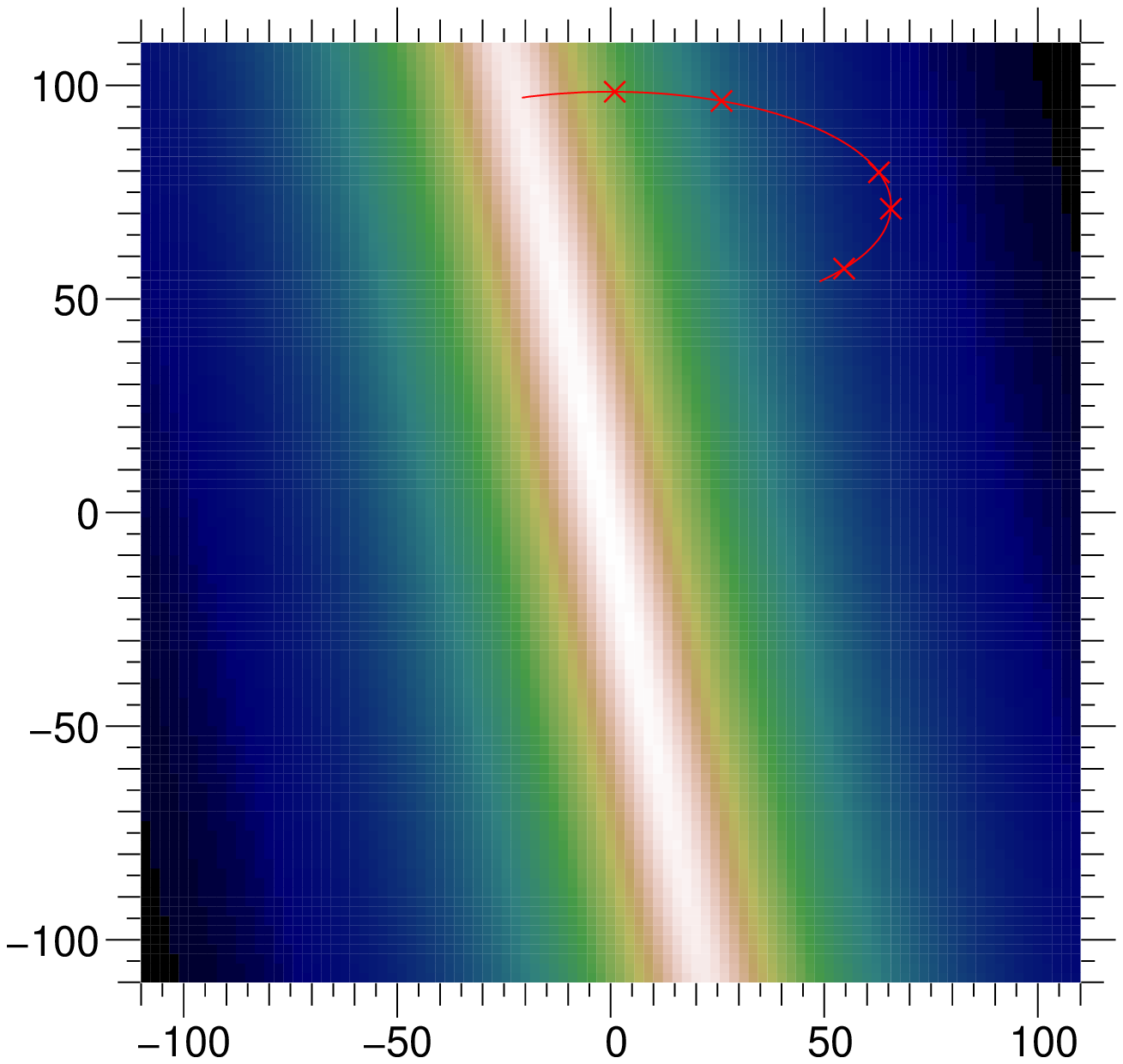}
\hspace{0.5cm}
\includegraphics[height=6cm]{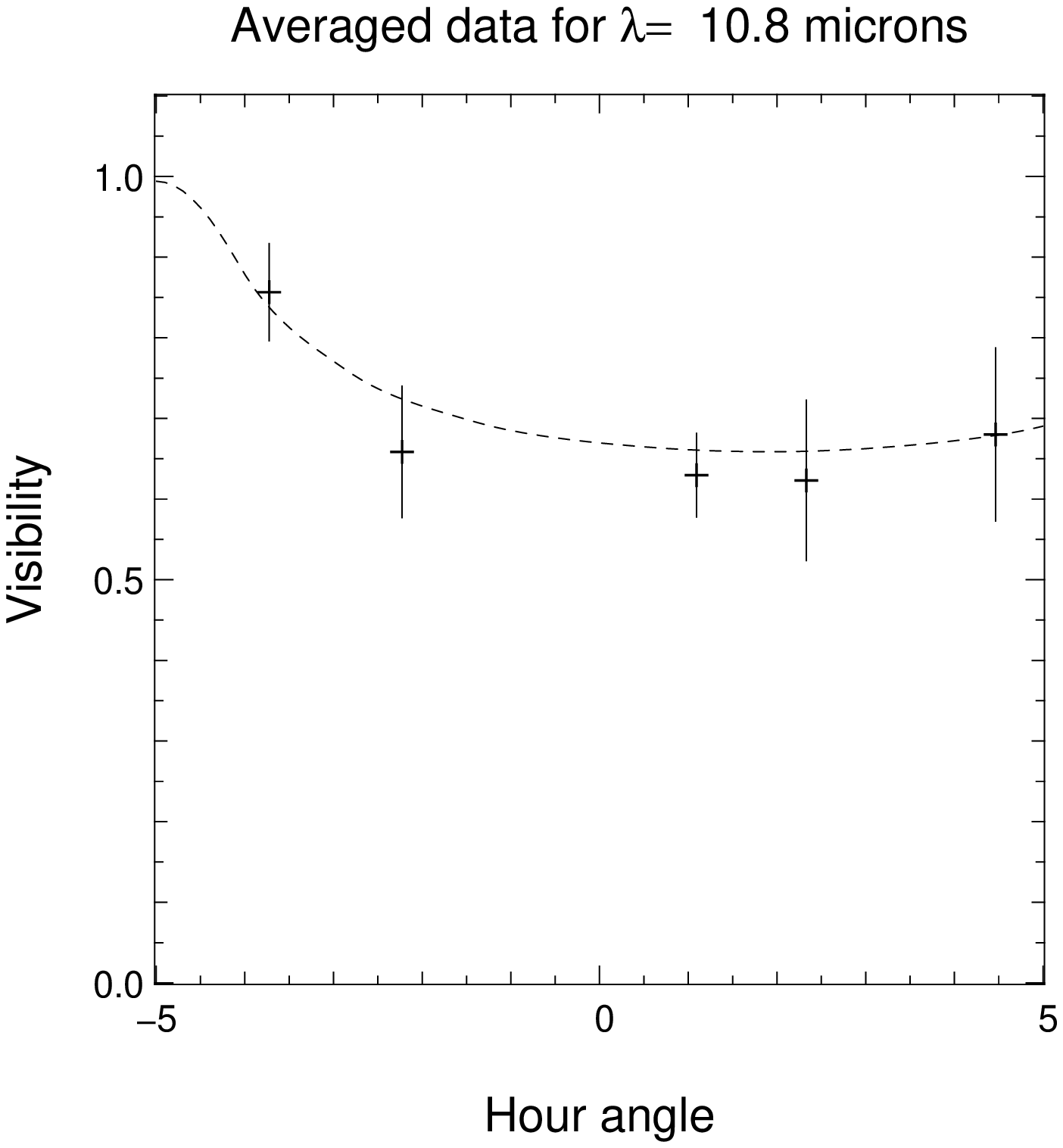}} 
\caption{{\bf Left:}51 Oph disk visibility model. The five visibility points (marked as crosses on the $uv$-track) were obtained in 2 different nights. {\bf Right:} Visibilities as a function of hour angle. Best fit for a standard disk with an inclination of 88$^\circ$ and a position angle of 78$^\circ$.\protect\cite{gil:uv}}
\end{figure*}

In Figure~\ref{gil:uv} on the left, we can see the visibility model for the 51 Oph circumstellar disk and the $uv$-track for our measurements. The 5 visibility points are marked with crosses over the $uv$-track. In this figure north is up and west is left, so we note that the position angle is obtained after a Fourier transform of the visibility model (-12$^\circ$+90$^\circ$=78$^\circ$). We have also studied the visibilities behavior against projected baselines and hour angle. In Figure~\ref{gil:uv} on the right, we have plotted the visibilities as a function of hour angle. For both cases we have found a good fit for all the visibility points by using the disk model parameters in Table~\ref{gil:model}.

\section{Discussion}

The MIDI observations are compatible with a flat circumstellar disk, inclined of 88$^\circ$ with a position angle of 78$^\circ$. These results are compatible with the recent results obtained by Thi et al. \cite{gil:Thi}, who have found a high column density of CO rotating in a disk in the inner astronomical unit of 51 Oph, seen nearly edge-on. 
We have determined a stellar radius of 7 $R_{\odot}$, a disk inner radius of 0.55 AU, an outer radius of 7 AU, a disk inclination angle of about 88$^\circ$ and a disk position angle of about 78$^\circ$. Due to the nature of the object, our interpretation relies critically on one visibility point and thus drew some criticism. Four of the visibility points are nearly identical, while the fifth one is significantly higher. This higher visibility point was obtained with the lowest projected baseline at 85.6m. There is consequently a potential degeneracy of information between a base length effect and an angle effect. In order to secure our data set, we plan to obtain a few extra visibility points with MIDI.
 
\section{Conclusions}
\begin{itemize}
      \item 51 Oph was observed for the first time in the mid-infrared at high-angular resolution.
      \item We have modeled 51 Oph visibilities and were able to constrain the size and geometry of the 51 Oph circumstellar disk.
      \item The best fit to the data corresponds to a flat circumstellar disk, inclined of 88$^\circ$ with a position angle of 78$^\circ$.
      \item We will need more visibility measurements in order to secure our data set and confirm the results obtained.
       \end{itemize}

%
%



\end{document}